\documentclass{emulateapj}
\usepackage{natbib}
\usepackage{graphicx}
\usepackage{amsmath}
\usepackage{epsfig}
\begin{document}
\title{Formaldehyde and methanol deuteration in protostars: fossiles
  from a past fast high density pre-collapse phase} 
\author{V. Taquet$^1$ , C. Ceccarelli$^1$ , C. Kahane$^{1}$}
\altaffiltext{1}{UJF-Grenoble 1 / CNRS-INSU, Institut de Plan\'{e}tologie et d\textquoteright Astrophysique de Grenoble (IPAG) UMR 5274, Grenoble, F-38041, France}

    \date{Received - ; accepted -}
\begin{abstract}
  Extremely high deuteration of several molecules have been observed
  around low mass protostars since a decade. Among them, formaldehyde
  and methanol present particularly high deuteration, with
  observations of abundant doubly and triply deuterated forms. Both
  species are thought to be mainly formed on interstellar grains
  during the low temperature and dense pre-collapse phase by H and D
  atom additions on the iced CO. We present here a theoretical study
  of the formaldehyde and methanol deuteration obtained with our
  gas-grain model, GRAINOBLE. This model takes into account the
  multilayer nature of the mantle and explores the robustness of the
  results against the uncertainties of poorly constrained chemical and
  surface model parameters.  The comparison of the model predictions
  with the observations leads to two major results: i) the observed high deuteration is
  obtained during the last phase of the pre-collapse stage, when the density
  reaches $\sim 5\times 10^6$ cm$^{-3}$, and this phase is fast,
  lasting only several thousands years. ii) D and H
  abstraction and substitution reactions are crucial in making up the
  observed deuteration ratios; This work shows the power of
  chemical composition as a tool to reconstruct the past
  history of protostars.
 \end{abstract}


\section{Introduction}

Although the deuterium elemental abundance is $1.5 \times 10^{-5}$
relative to hydrogen \citep{Linsky2003}, observations carried out
during the last decade have revealed high abundances of singly,
doubly, and even triply deuterated molecules in low mass pre-stellar
cores and Class 0 protostars \citep[][]{Ceccarelli2007}. In this work, we focus on formaldehyde and
methanol, both species likely synthesized on the grain surfaces
{\citep{Tielens1982, Watanabe2002}} {as gas phase
reactions are unable to reproduce the large observed abundances \citep{Roberts2004, Maret2005}}.

In pre-stellar cores, HDCO/H$_2$CO and D$_2$CO/H$_2$CO abundance
ratios up to 10\% have been observed {\citep{Bacmann2003, Bergman2011}}. The observed
deuterium fractionations increase with the increasing CO depletion,
suggesting that the latter is a key parameter. Analogously, several formaldehyde and methanol
isotopologues have been detected in Class 0 low mass protostars
\citep{Ceccarelli1998,Ceccarelli2001, Parise2002,Parise2004,Parise2006}. 
However, unlike in pre-stellar cores, no
correlation between the CO depletion and deuterium fractionation is
observed, so that the deuteration process is thought to occur in the previous
pre-collapse phase.  In general, methanol is more enriched in
deuterium than formaldehyde. In most of the observed sources, the
simply deuterated molecule HDCO shows a deuterium fractionation
between 13 and 20\% whereas CH$_2$DOH shows a fractionation between 37
and 65\% \citep{Parise2006}. CHD$_2$OH also shows a higher
fractionation than D$_2$CO, by a factor of $\sim 2$.  Finally, the
deuteration process occurs more efficiently on the methyl group than
on the hydroxyl group of methanol, as the [CH$_2$DOH]/[CH$_3$OD] ratio
has been observed to be between 10 and 20 in a sample of low- and
intermediate- mass protostars by \citet{Parise2006} and \citet{Ratajczak2011}.

Recent experimental works have confirmed
the synthesis of formaldehyde and methanol by hydrogenation of iced CO
{\citep{Watanabe2002,Hidaka2007, Fuchs2009}} and also highlighted the complex
chemical evolution of H$_2$O-CO ices when they are exposed to D and H
atoms. Indeed, \citet{Nagaoka2005}, \citet{Nagaoka2007} and
\citet{Hidaka2009} have shown that formaldehyde and methanol can be
efficiently deuterated into HDCO, D$_2$CO, CH$_2$DOH, CHD$_2$OH and
CD$_3$OH, when they are irradiated by D atoms. {Conversely, abstractions 
of D can only occur on formaldehyde, and not on methanol. Therefore, 
formaldehyde isotopologues only can be hydrogenated to form back HDCO 
and H$_2$CO if they are exposed to H atoms.  }
The relative reaction rates deduced from these works show that
H and D abstractions and substitutions on formaldehyde and methanol
are as efficient as addition reactions \citep[see also the theoretical
calculations by][]{Goumans2011a,Goumans2011b}. 
These processes could, therefore, largely
increase the deuterium fractionations of formaldehyde and methanol
after the complete depletion of CO and their formation on the
surfaces.

In the past, astrochemical models have struggled to reproduce the
observed deuteration ratios of formaldehyde and methanol. While it is
now clear that CO depletion plays a major role in increasing the
atomic D/H ratio of the gas landing on the grain surfaces
\citep{Roberts2004}, a full model coupling the gas and
grain chemistry that {\it simultaneously} reproduces the observed
formaldehyde and methanol deuteration is still missing. A previous
{attempt} was carried out by \citet{Caselli2002} and
\citet{Stantcheva2003}, who studied the formaldehyde and methanol
deuteration as function of the atomic D/H ratio, taken as a free
parameter. However, as also emphasised by \citet{Parise2006}, the
use of a constant [D]/[H] ratio is i) unable to predict all the
deuterium fractionations at the same time and ii) is not necessarily
correct, as formaldehyde and methanol may be formed on the grains at
different times. {\citet{Cazaux2011} have studied the deuteration 
of water and formaldehyde (and not methanol) by coupling gas phase
and grain mantle chemical networks.}

In this article, we re-consider the problem of the formaldehyde and
methanol deuteration, using our gas-grain coupled model,
GRAINOBLE, that takes into account D and H atoms addition, but also
abstraction and substitution reactions. 
{Our goal is to simultaneously reproduce the formaldehyde and 
methanol deuteration observed towards Class 0 protostars by 
considering their formation on interstellar grains only. Note that, 
once sublimated from ices, the abundance of these species is only 
slightly affected by gas phase reactions, as the typical chemical 
timescale \citep[$\sim 10^5$ yr, ][]{Charnley1997} is larger than 
the typical age of these objects \citep[$\sim 10^4$ yr, ][]{Andre2000}.}
%
We will show that understanding how and when the observed deuteration occurs
will also provide us with hints on the process itself and on the past
history of the protostars.

\section{Multilayer modeling of deuterated ices} \label{modeling}

\subsection{Description of the GRAINOBLE model}

For this study we use the GRAINOBLE model \citep[][hereafter TCK11]{Taquet2011}.
 Briefly, it is a gas-grain chemical model, based on
the rate equations approach introduced by \citet{Hasegawa1992} for
time-dependent grain surface chemistry modeling. It considers the
following four processes:\\
i) Accretion of gas phase species onto the grain surfaces as function of time.\\
ii) Thermal diffusion of adsorbed species. The hopping
rate follows a Boltzmann function which depends on the
diffusion energy $E_d$.\\
iii) Surface reactions via the Langmuir-Hinshelwood mechanism. The
probability of the reaction is given by the exponential portion of the quantum mechanical probability for tunneling through a square barrier and therefore depends on the activation energy $E_a$ of
the reaction.\\
iv) Thermal evaporation caused by the thermal balance and by
the cosmic-ray induced heating of the grains \citep{Hasegawa1993a}. 
{We ignored the photolytic process due to the cosmic-ray induced UV photons because
they have a negligible influence on the abundance of formaldehyde and methanol in dark clouds
\citep[5\% at maximum, ][]{Cuppen2009}.}

We follow the mantle formation on grains with a multilayer approach in
which the outermost layer only is reactive, while the mantle bulk
remains inert. Unlike \citet{Hasegawa1993b} and \citet{Garrod2011}
models, the trapping of particles into the bulk is performed one layer
at a time, once the considered layer is filled.

\subsection{Chemical network} 

{We consider the accretion of gaseous H, D, H$_2$, HD, D$_2$, O, and CO onto grains.}

We assume that formaldehyde, methanol and their
deuterated isotopologues are formed via hydrogenation and deuteration
addition reactions on molecules initiated by the accretion of
CO. Based on experimental and theoretical works \citep{Hidaka2007, Watanabe2008, 
Andersson2011}, the energy barriers of the D and H addition
reactions {involving CO and formaldehyde} are very similar, so we 
assume them to be identical {($=E_a$)}. Note that, however, given the high uncertainty in the value of $E_a$ (see
TCK11), this is taken as a free parameter. 
{In contrast, reactions involving a radical are barrierless.}

\begin{figure*}[htp]
\centering
\includegraphics[width=180mm]{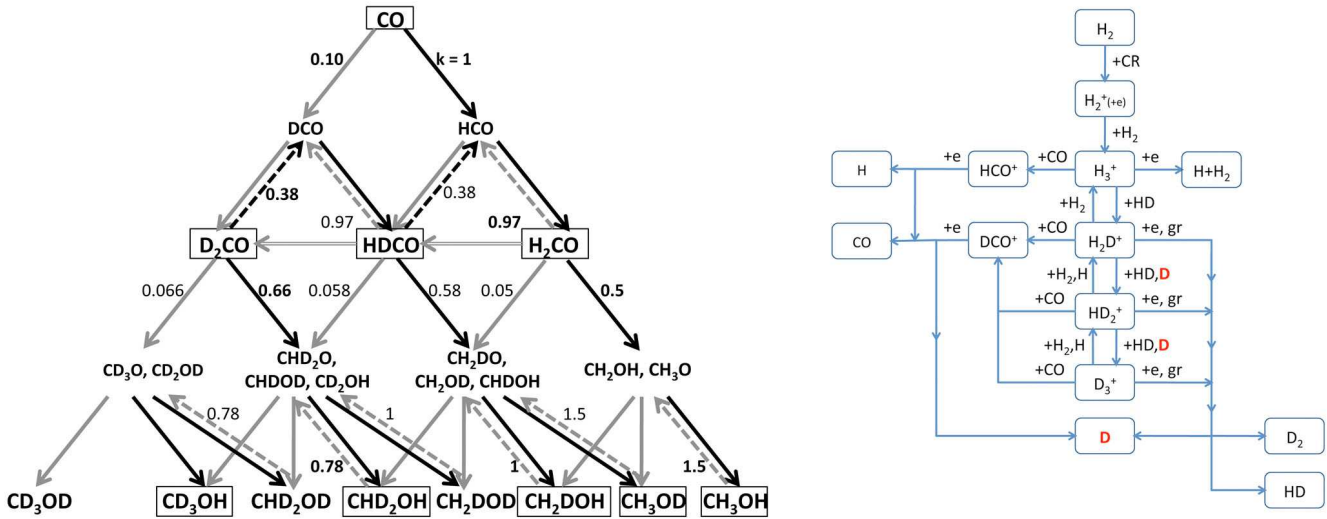}
\caption{\textit{left)} chemical network on grain mantles focusing on the formation
  of formaldehyde and methanol. Solid, dashed, and open arrows refer to addition, 
  abstraction, and substitution reactions. Black arrows refer to 
  reactions involving H while grey arrows represent reactions 
  involving D. Finally, the bold values refer to the rates relative 
  to the H + CO reaction experimentally deduced by \citet{Nagaoka2007} and
   \citet{Hidaka2009} while non-bold values show the deduced relative rates 
   by analogy with similar reactions.
  \textit{right)} Chemical network in the gas phase leading to
  atomic deuterium formation.}
\label{networks}
\end{figure*}

Following the experimental works of \citet{Hidaka2009} and
\citet{Nagaoka2007}, we also include the abstraction and substitution
reactions of H and D on formaldehyde and methanol according the scheme
proposed by \citet{Watanabe2008} and \citet{Hidaka2009}, and shown in
Fig. \ref{networks}.  We adopt the probability of each reaction following the 
relative rates deduced experimentally when they are available, as marked in the Figure. 
For the reactions that have not been derived by
the experiments, we adopt the probability measured for the same
isotopologue by analogy. For example, the reaction D$_2$CO + D
$\rightarrow$ CD$_3$O has a reaction rate equal to the H + CO rate
multiplied by 0.1 (D + CO) and 0.66 (D$_2$CO + H).

Even if most of water is likely formed during the translucent phase, and therefore 
before H$_2$CO and CH$_3$OH \citep[see][]{Oberg2011}, our model also takes into account its formation, 
as it is in competition with formaldehyde and methanol formation according to TCK11.
With respect to TCK11, we added the deuteration reactions and the path \\
\begin{center} H$_2$ + OH $\rightarrow$  H$_2$O + H \end{center}
which seems to be the most efficient reaction in molecular clouds, as suggested by
\citet{Cuppen2007}.
 
Since the deuteration of formaldehyde and methanol on the grains
depends on the atomic gas phase D/H ratio, we are here particularly
interested in the chemical network leading to the formation of atomic
deuterium in the gas phase. We use the fractionation reactions
introduced by \citet{Roberts2000, Roberts2004} to study the formation of H$_3^+$ isotopologues
and atomic deuterium. The gas phase chemical network is shown in
Figure \ref{networks}. {As shown by \citet{Flower2006}, the ortho/para ratio (opr)
of H$_2$ can influence the deuteration of H$_3^+$ and consequently the atomic D/H ratio, 
but only when it is $\geq 10^{-3}$. 
The available measures in cold gas indicate low H$_2$ opr values \citep[$< 10^{-3} - 10^{-2}$, ][and references therein]{Dislaire2012}. We, therefore, did not consider this effect in this work.}

\subsection{The physical model}

Rather than simulating the evolution from the diffuse cloud state to
the pre-stellar core, we focus on the last stage of the evolution, when
the material is already molecular and the density reaches the value
$n_H$. 
{The initial abundance of gas phase species (H, D, HD, D$_2$, H$_3^+$ 
isotopologues, ...) are given by the steady-state abundance, obtained 
solving the gas phase chemical network presented in \S 2.2
and considering the recombination of H and D onto grains. 
The abundance relative to H nuclei of the deuterium and CO reservoirs are
 $1.5 \times 10^{-5}$ and $4.75 \times 10^{-5}$.}

We follow, then, the formation and evolution of the grain
mantles during this phase, keeping the gas density $n_H$
constant. Briefly, as the time passes, CO freezes out onto the grain
mantles, where it forms H$_2$CO and CH$_3$OH and their deuterated
isotopologues. Therefore, the deuteration of the formaldehyde and methanol on the mantle primarily
depends on the gaseous atomic D/H ratio.

\subsection{Multi-parameter approach}

We use a multi-parameter approach as described in TCK11, to
take into account the variation or the uncertainty of physical
conditions and surface parameters.

Low mass pre-stellar cores are the place where most of formaldehyde
and methanol are assumed to be formed. These objects show spatial
distributions of densities of H nuclei between $10^4$ and $5\times 10^6$
cm$^{-3}$ and temperatures between 8 and 12 K
\citep[see][]{Crapsi2007}. Accordingly, we consider here
four density values ($n_H = 10^4$, $10^5$, $10^6$, $5\times 10^6$
cm$^{-3}$) and three temperatures ($T_g = T_d = 8, 10, 12$ K).
Besides, prestellar cores show an increase of the grain sizes caused
by the coagulation process whose the efficiency increases with the
density \citep{Flower2005,Vastel2006}. We adopt three grain sizes
$a_d = 0.1, 0.2, 0.3$ $\mu$m, which affect the depletion rates and
therefore the density of H and D in the gas phase.

As discussed in TCK11, the diffusion to desorption energy $E_d/E_b$
ratio mainly depends on the ice properties. We, therefore, vary
$E_d/E_b$ between 0.5 and 0.8, as suggested by experimental studies.
The activation energy $E_a$ of the reactions {involving CO and formaldehyde}
is also a free parameter. By comparing the model predictions with the observations of solid CO and
methanol, we have deduced in TCK11 that $E_a$ must be lower than 1500
K, so that we restrain the value of $E_a$ between 400 and 1400 K.  
Light particles (H, D, H$_2$, HD, D$_2$) that accrete onto the ices
have a binding energy distribution that depends on the ice
properties, and the adsorption conditions of the
adsorbate \citep[see for example][]{Hornekaer2005}.
We therefore consider the binding energies relative to amorphous water ice 
of these light species as a free parameter whose values are: 400, 500, 600 K.  

As the abundance of the gaseous atomic oxygen in the pre-stellar cores
is uncertain, we consider three values of oxygen abundance relative to
H nuclei $X($O)$_{ini}$: $10^{-8}$, $10^{-6}$, $10^{-4}$. 
 
Finally, in TCK11, we have shown that the site size $d_s$ and the porosity
factor $F_{por}$ have a little impact on the chemical composition of
the grain mantles, so we assume $d_s = 3.1$ $\AA$ 
\citep{Jenniskens1995}, and $F_{por} = 0$.

We run a grid of 2916 models in which we vary the parameters described
above and listed in Table \ref{table_grid}. For each density $n_H$, we compute the mean
fractionation of each isotopologue with its {1 sigma} standard deviation, following \citet{Wakelam2010}.

\begin{table}[htp]
\centering
\caption{List of parameters and the values range explored in
  this work.}
\begin{tabular}{c c}
\hline
\hline
Parameter & Values  \\
\hline
Density $n_H$ & $10^{4}$ - ${10^5}$ - $10^{6}$ - $5 \times 10^{6}$ cm$^{-3}$ \\
Temperature $T_g = T_d$ & 8 - {10} - 12 K \\
Grain size $a_{d}$ & 0.1 - {0.2} - 0.3 $\mu$m \\
Energy ratio $E_d/E_b$ & 0.5 - {0.65} - 0.8 \\
Binding energy $E_{b,wat}$(H) & 400 - {500} - 600 K \\
Activation energy $E_a$ & 400 - {900} - 1400 K \\
Initial oxygen abundance $X$(O)$_{ini}$ & $10^{-8}$ - ${10^{-6}}$ - $10^{-4}$ \\
\hline
\end{tabular}
\label{table_grid}
\end{table}

 \section{Results}
 
\subsection{The D/H gas phase ratio}

At first approximation, the steady state densities of H and D, assumed
as initial conditions, are roughly constant regardless the total
density and their abundance relative to H nuclei, therefore, decrease with
increasing $n_H$. 
At low densities, a significant fraction ($\sim$30\%)
of the deuterium reservoir is already in atomic form before the depletion of CO: 
the increase of the atomic D/H ratio with the CO depletion will therefore be low. 
On the contrary, at high densities, only a negligible fraction of
deuterium is in the atomic form at the beginning, so that the atomic
D/H ratio strongly increases with the CO depletion. The
larger the density, the larger the gaseous atomic D/H ratio increase
with the CO depletion, as highlighted in Figure
\ref{comp_ref_addabst_dXmantle_Dratio_COdepl_multi} which shows the
evolution of the atomic D/H as function of CO depletion, for $n_H$
equal to $10^5$ and $5\times10^6$ cm$^{-3}$ respectively.

 \subsection{Model with addition reactions only}
 
\begin{figure}[htp]
\centering
\includegraphics[width=88mm]{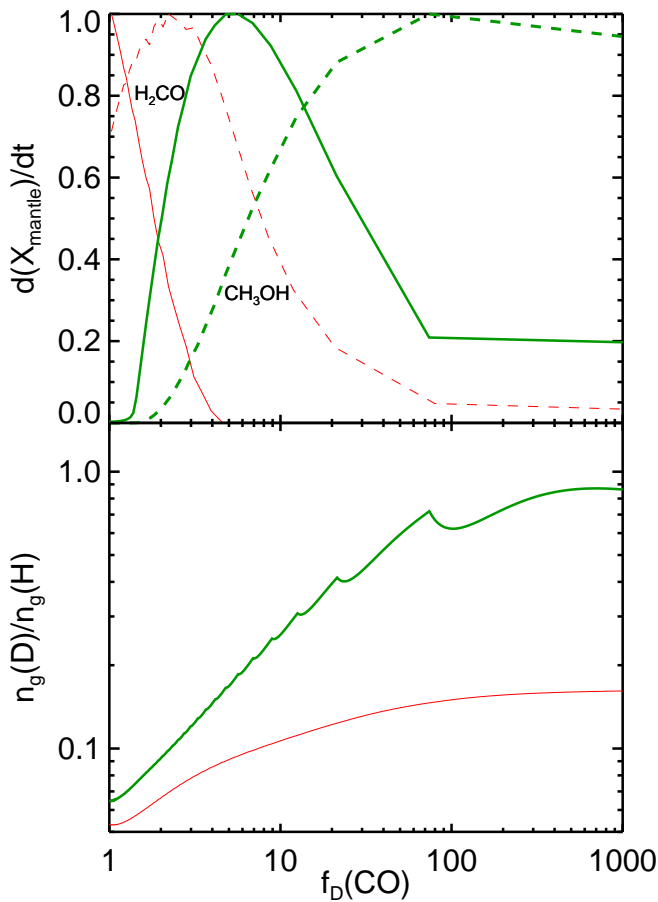}
\caption{Normalized H$_2$CO (solid lines) and CH$_3$OH (dashed lines)
  formation rates (top panel) and gaseous atomic D/H ratio (bottom panel)
  as function of the CO depletion factor $f_D$(CO) =
  $n_{g}(\textrm{CO})/n_{g,ini}(\textrm{CO})$, for $n_H = 10^5$ (thin
  red lines), and $5\times 10^6$ (thick green lines) cm$^{-3}$. 
  {The spikes are caused by the multilayer nature of the mantle.}}
\label{comp_ref_addabst_dXmantle_Dratio_COdepl_multi}
\end{figure}

 In this section, only addition reactions are considered on grain
 mantles (see \S 2.2).  In this case, the deuteration of formaldehyde
 and methanol primarily depends on two factors, as illustrated in
 Fig. \ref{comp_ref_addabst_dXmantle_Dratio_COdepl_multi}: i)
 \textit{The increase of the gas phase atomic D/H ratio with the CO
   depletion:} the ratio increases with increasing CO depletion,
 namely with time, and with increasing density (see above); ii)
 \textit{When formaldehyde and methanol are formed:} as explained in
 detail in TCK11, the increase of $n_H$ delays the formation of
 formaldehyde and methanol. In general, the two effects results in 
 a larger deuteration of formaldehyde and methanol for
 larger densities and larger evolutionary times.

\begin{figure}[htp]
\centering
\includegraphics[width=70mm]{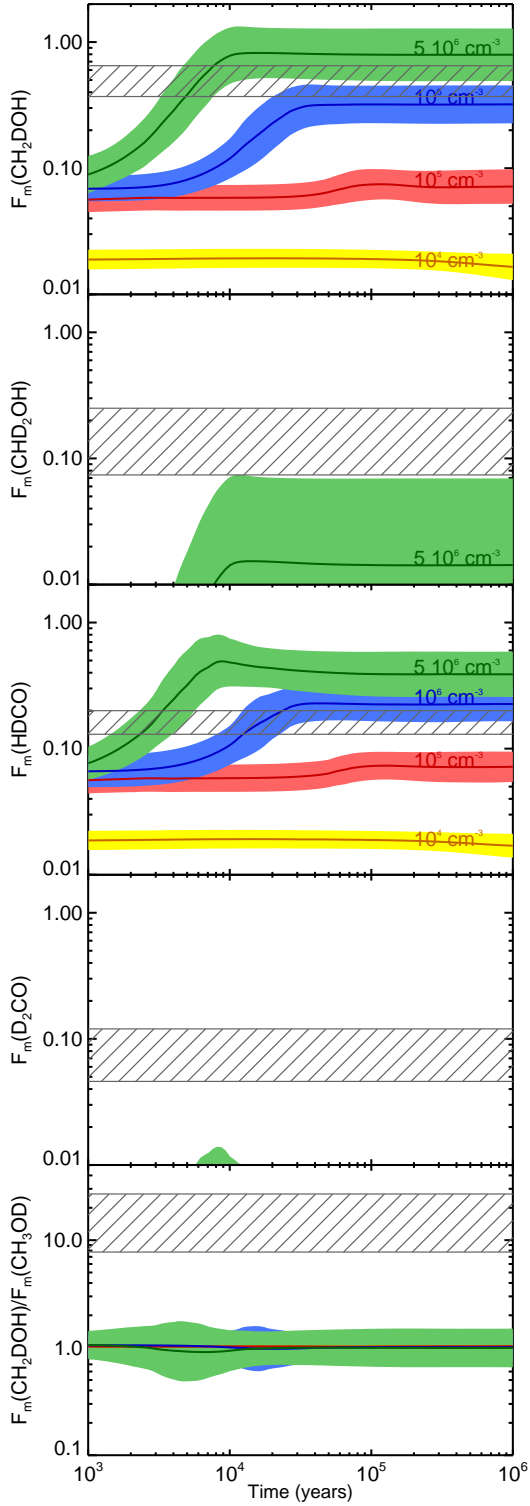}
\caption{Mean deuterium fractionation of methanol and
  formaldehyde (solid lines) with their {1 sigma} standard deviation (color
  levels) with time, obtained considering addition reactions only (\S
  3.2). The hatched zones give the range of observed values
  \citep{Parise2006}.}
\label{Fmantle_time_all_nH_2_withoutabstract}
\end{figure}

This is illustrated in
Fig. \ref{Fmantle_time_all_nH_2_withoutabstract}, which shows the
temporal evolution of the mean deuterium fractionations of iced
formaldehyde and methanol along with their uncertainty, for different
densities. The high density cases ($n_H \geq 10^6$ cm$^{-3}$) show
large enhancements of the CH$_2$DOH/CH$_3$OH and HDCO/H$_2$CO ratios
with time, reaching the unity for the singly deuterated forms. No
increase of the deuterium fractionation is, on the contrary, observed
at low densities ($n_H = 10^4 - 10^5$ cm$^{-3}$), the deuteration
ratios never exceeding 0.02.
On the same figure, we report the range of observed values \citep[from][]{Parise2006}. 
No model can reproduce the full set of
observations. Indeed, the high density cases predict
CH$_2$DOH/CH$_3$OH and HDCO/H$_2$CO ratios in agreement with
observations for times between $5\times 10^3$ and $10^6$ yr. However,
they cannot predict the observed high abundances of the doubly
deuterated molecules. In addition, observations show that methanol is
three times more deuterated than formaldehyde whereas the models of
Fig. \ref{Fmantle_time_all_nH_2_withoutabstract} predict a factor 1.5
at most. We conclude that this class of models fails to reproduce the
observations.

\subsection{Abstraction and substitution reactions} \label{sec:results_abstract}

\begin{figure}[htp]
\centering
\includegraphics[width=70mm]{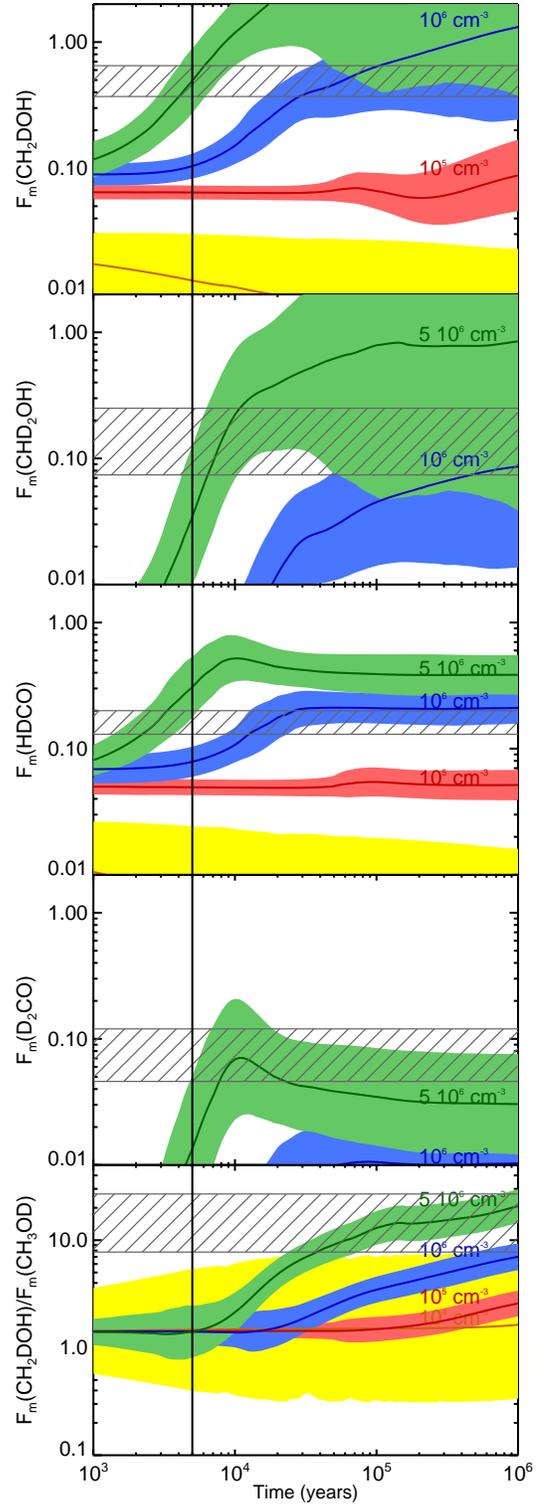}
\caption{Mean deuterium fractionation of methanol and
  formaldehyde (solid lines) with their {1 sigma} standard deviation (color
  levels) with time, obtained adding the abstraction and
  substitution reactions (\S 3.3). The hatched zones give the range
  of observed values \citep{Parise2006}.}
\label{Fmantle_time_all_nH_2_withabstract}
\end{figure}

Figure \ref{Fmantle_time_all_nH_2_withabstract} shows the temporal
evolutions of formaldehyde and methanol deuterium fractionations when 
 abstraction and substitution reactions are included. From
Fig. \ref{Fmantle_time_all_nH_2_withabstract} it is clear that the
inclusion of the addition and substitution reactions strongly increase
the deuterium fractionations, especially at high densities.  For
densities $\sim 5 \times 10^6$ cm$^{-3}$, the model predicts
fractionations of doubly-deuterated molecules compatible with
observations, at time $\sim 5\times 10^3$ yr. Unlike CH$_2$DOH, which
shows an increase of deuteration compared to
Fig. \ref{Fmantle_time_all_nH_2_withoutabstract} and more particularly
at longer timescales, the HDCO deuteration is not enhanced. This is
due to the efficiency of deuterium abstraction on HDCO which allows
the formation of H$_2$CO whereas deuterium abstraction reactions on
CH$_2$DOH, leading to CH$_3$OH formation, are negligible. The observed
[CH$_2$DOH]/[HDCO] ratio of 3 can now be predicted. Finally, because
the abstraction reactions can only occur significantly on the methyl
group of methanol and not on its hydroxyl group, the
[CH$_2$DOH]/[CH$_3$OD] ratio in enhanced. The observed ratio of 10-20
can also be predicted but at a larger time, $5 - 10 \times 10^4$ yr.
We conclude that this class of models succeeds to reproduce all the
observations simultaneously, with the exception of the
[CH$_2$DOH]/[CH$_3$OD] ratio, for a density of about $5 \times 10^6$
cm$^{-3}$ and at a time of 5000 yr.

\section{Discussion and conclusions}

Although multiply deuterated molecules in low mass protostars have
been discovered and observed for a decade, published models have had
difficulties in reproducing the observed abundances, especially those
of formaldehyde and methanol \citep{Parise2006}. Both species are
believed to be mainly synthesised and to be observed on the grain surfaces during the cold and
dense pre-collapse phase \citep[see][]{Oberg2011} and that they are observed in the gas when
they sublimate off the grain mantles upon heating from the central
star. On the grains, H$_2$CO and CH$_3$OH are thought to be the result
of the hydrogenation of iced CO. Their deuteration, therefore, depends
on when exactly the two species are formed and how. We have shown
that, based on our grain surface model GRAINOBLE, formation of H$_2$CO 
and CH$_3$OH from addition reactions alone fails to predict the observed 
deuterium fractionation. In contrast, if D and H abstraction and substitution
reactions are added, the GRAINOBLE model can reproduce {\it
  simultaneously} the observed values. Therefore, these processes are crucial and
more laboratory experiments and theoretical computations are needed to
better constrain their rates on the ices. Reproducing the observed [CH$_2$DOH]/[CH$_3$OD]
ratio remains a challenge, as it is still underestimated by the
model. Previous studies have
suggested that this may be caused by D and H exchanges on the ices
during the sublimation phase \citep{Ratajczak2009} or activated by
photolysis processes \citep{Weber2009} or on the gas phase \citep{Osamura2004},
 all processes that would be inefficient in altering the
other isotopologues \citep[see][for a detailed
discussion]{Ratajczak2011}.

With the above exception, our model predicts the observed abundance
ratios for high densities ($\sim 5\times 10^6$ cm$^{-3}$) and for
relatively short times ($\sim 5000$ yr).  We emphasise that this time
corresponds only to the final stage at high density and not to the
age of the condensation, which can be considerably larger. In fact, it
is possible and even likely that the pre-stellar cores spend a long
time in a less dense phase \citep[e.g. ][]{Bergin2007}. 
However, the comparison of the observed H$_2$CO and CH$_3$OH
deuteration with our model predictions suggests that the last phase at
high density is short, just a few thousands years. In other
words, soon after the central density reaches $\sim 5\times 10^6$
cm$^{-3}$ the {protostellar collapse} starts.  This is a nice illustration
of how chemistry can help us to understand the past history of the
protostar.

In conclusion, our work leads to two important results:\\
1) The pre-collapse phase may last hundred thousands of years:
however, when the central density reaches $\sim 5\times 10^6$
cm$^{-3}$, the collapse starts in about a few thousands years.
2) D and H atoms abstraction and substitution reactions are crucial
in the grain surface chemistry and should be incorporated into models. 
{Thus, more experimental and theoretical works are therefore needed to better contrain
their efficiency and therefore the timescale needed to reproduce the observations.} \\
 
\begin{acknowledgements}

  This work has been supported by l\textquoteright Agence Nationale
  pour la Recherche (ANR), France (project FORCOMS, contracts
  ANR-08-BLAN-022).

\end{acknowledgements}

\end{document}